\documentclass[aps,prd,10pt,superscriptaddress,onecolumn,nofootinbib,preprint]{revtex4-1}
\usepackage{graphicx, epsfig} 
\usepackage{amsmath,amssymb,amsfonts,dsfont,mathrsfs,amsthm,mathtools}
\usepackage{bm} 
\usepackage{color}
\usepackage{physics}
\usepackage[usenames]{xcolor}
\usepackage{hyperref}
\usepackage{siunitx}
\hypersetup{colorlinks=true,urlcolor=blue,linkcolor=magenta,citecolor=blue,filecolor=blue}
\usepackage[normalem]{ulem}
\usepackage{array}
\usepackage{hyperref}
\usepackage{booktabs}
\usepackage{cancel}
\usepackage{blindtext}
\usepackage{tikzsymbols} 
\usepackage[paperwidth=625pt,top=80pt,right=60pt,bottom=80pt,left=60pt,headheight=50pt]{geometry}
\usepackage{lipsum}

\newcommand{\diff}[1]{\text{d}#1}

\begin{document}

\title{Nonconformally Ricci-flat instantons in Conformal Gravity with and without nonlinear matter fields}
\begin{abstract}
In this work, we study nonconformally Ricci-flat gravitational instantons in four-dimensional Conformal Gravity, both in vacuum and in the presence of nonlinear conformal matter. First, the one-parameter extension of the Kerr-NUT-AdS metric is analyzed. We obtain their conserved charges by using the Noether-Wald formalism. It turns out that they receive corrections from the linear modes present in Conformal Gravity, which are properly identified. Then, we perform the analytic continuation into the Euclidean section and find the curve in parameter space along which this solution becomes regular and globally (anti)self-dual. Using the Dunajski-Tod theorem, we show that the self-dual metric is not conformally Ricci-flat. Then, the backreaction of nonlinear conformal matter is considered. In particular, we find new gravitational instantons in the presence of conformally coupled scalar fields and ModMax electrodynamics. We compute the partition function and conserved charges, which turn out to be finite by virtue of the conformal invariance of the theory. As a byproduct, we also obtain a generalization of the Riegert metric dressed with nonlinear conformal matter as a particular limit of these instantons. For all cases, we analyze the global properties, the curve in parameter space where the solutions are (anti)-self-dual, and the on-shell Euclidean action, among other features.
\end{abstract}

\author{Crist\'obal \surname{Corral}}
\email{cristobal.corral@uai.cl}
\affiliation{Departamento de Ciencias, Facultad de Artes Liberales, Universidad Adolfo Ib\'{a}\~{n}ez, Avenida Padre Hurtado 750, 2562340, Vi\~{n}a del Mar, Chile}

\author{Borja \surname{Diez}}
\email{borjadiez1014@gmail.com}
\affiliation{Departamento de Ciencias, Facultad de Artes Liberales, Universidad Adolfo Ib\'{a}\~{n}ez, Avenida Padre Hurtado 750, 2562340, Vi\~{n}a del Mar, Chile}

\author{Eleftherios \surname{Papantonopoulos}}
\email{lpapa@central.ntua.gr}
\affiliation{Physics Division, School of Applied Mathematical and Physical Sciences, National Technical University of Athens, 15780 Zografou Campus, Athens, Greece.}

\maketitle

\section{Introduction}

The anti–de Sitter/conformal field theory (AdS/CFT) correspondence has provided a powerful tool for studying strongly interacting field theories at the fixed points of the renormalization group flow, where standard perturbative techniques are no longer valid~\cite{Maldacena:1997re,Witten:1998qj,Gubser:1998bc}. At those points, quantum field theories exhibit conformal invariance, which dictates the specific form of correlator functions of the theory. In the context of gravity, theories that possess conformal symmetry display an improved ultraviolet behavior, providing a suitable scenario for studying quantum effects. Conformal gravity, for instance, is a particular example of the latter. It appears as a natural ultraviolet completion of general relativity, although its symmetry becomes anomalous at one-loop level~\cite{Adler:1982ri} (for a review of Conformal Gravity see~\cite{Mannheim:2005bfa,Mannheim:2016lnx}). In asymptotically locally AdS spacetimes, Grumiller et al. have shown that Conformal Gravity is finite without introducing boundary counterterms~\cite{Grumiller:2013mxa}. This theory has a well-posed variational principle for the holographic sources, i.e. the boundary metric and the partially massless response, rendering a rich scenario for studying holography~\cite{Grumiller:2013mxa}. Witten and Berkovits, on the other hand, showed that the supersymmetric extension of Conformal Gravity appears naturally in string theory on twistor spaces~\cite{Berkovits:2004jj}. Furthermore, it provides a natural embedding for renormalizing Einstein gravity on asymptotically locally AdS spacetimes when Neumann boundary conditions on the Fefferman-Graham expansion are imposed~\cite{Maldacena:2011mk,Anastasiou:2016jix,Anastasiou:2020mik}. These boundary conditions eliminate ghost instabilities that appear due to the fourth-order nature of its field equations, rendering the theory free of pathologies~\cite{Lu:2011ks,Hell:2023rbf}.

Among the different configurations of Conformal Gravity~\cite{Riegert:1984zz,Mannheim:1988dj}, gravitational instantons are particularly relevant, as they allow one to study nonperturbative effects in quantum gravity. For instance, in Ref.~\cite{Strominger:1984zy}, Strominger, Horowitz, and Myers studied asymptotically flat gravitational instantons, representing contributions to the vacuum persistence amplitude between topologically inequivalent vacua. They computed the one-loop determinant in terms of the renormalization group of invariant masses and topological properties of instantons. Since the Conformal Gravity action is a quadratic form constructed out of the Weyl tensor, they show that (anti)self-dual instantons saturate a BPS bound, given in terms of the Chern-Pontryagin index---see Eq.~\eqref{BPSbound} below. More recently, extensions of the Euclidean Taub-NUT-AdS and Eguchi-Hanson metrics have been studied in Ref.~\cite{Corral:2021xsu}, where their conserved charges and partition function were computed. The latter are finite due to conformal invariance without invoking boundary terms. Furthermore, these configurations were shown to be inequivalent to the generalized Taub-NUT metrics studied in Refs.~\cite{Iwai:1992zu,Miyake:1995}. 

Although gravitational instantons in Conformal Gravity have been studied mainly in vacuum, therefore being Bach-flat, the presence of conformal matter backreacting on these topologically nontrivial configurations remains unexplored. In order to give a step toward this direction, one must consider a matter sector that preserves the symmetries of the gravity action. In four dimensions, this holds true for the Maxwell and Yang-Mills theories. However, this is not exclusive to gauge theories: conformally-coupled scalars can also be included without spoiling conformal invariance~\cite{Charmousis:2009cm}. Indeed, in Conformal Gravity theories, the presence of a scalar field that backreacts on the metric allows for the generation of hairy black hole solutions, while other configurations can be found via conformal mappings~\cite{Maeda:1988ab}. In all black hole solutions in the Einstein frame, the scalar field is coupled minimally to gravity. Applying a conformal transformation, other solutions can be obtained in the Jordan frame, which are not physically equivalent to the untransformed ones~\cite{Faraoni:1998qx}.

To find electrically and magnetically charged gravitational instantons, one must include backreacting Abelian gauge fields. However, when quantum effects are taken into account in any interacting gauge theory, their low-energy limit is described in terms of effective field theories~\cite{Koutsoumbas:2018gbd,Diakonou:1990fe,Diakonou:1990kb}. The standard procedure for obtaining the latter is integrating out the heavy degrees of freedom in the path integral formalism. In the context of quantum electrodynamics, Euler and Heisenberg showed that quantum effects generate gauge-invariant nonlinear terms at one-loop level~\cite{Heisenberg:1936nmg}. These irrelevant operators in the infrared encode quantum effects, such as vacuum polarization and photon self-interaction, to first order in the fine structure constant. Abelian gauge theories possessing higher-order terms are generically referred to as nonlinear electrodynamics, and they are characterized by strong electromagnetic fields that reduce to Maxwell’s theory in the weak field limit. For instance, the motion of particles around a magnetically charged Euler-Heisenberg black hole with scalar hair was studied in~\cite{Theodosopoulos:2023ice}. Nonlinear electrodynamics exhibits novel features, including the absence of initial singularities, an upper limit on the electric field at the origin for pointlike particles, and a finite self-energy of charged particles~\cite{Novello:1999pg}, as shown by Born and Infeld in their seminal paper~\cite{Born:1934gh}. 

Nevertheless, introducing nonlinear effects through higher-order operators in Maxwell theory is subtle. This is because, in Born-Infeld and Euler-Heisenberg theories, nonlinear interactions introduce a fixed energy scale that breaks conformal invariance. Nevertheless, there exists a particular nonlinear electrodynamics theory that preserves the same symmetries as Maxwell theory, i.e. conformal and $SO(2)$ duality invariance, which has been dubbed ModMax theory~\cite{Bandos:2020jsw}. Different aspects of this theory ---including its backreacting effects--- have been explored~\cite{Flores-Alfonso:2020euz,Flores-Alfonso:2020nnd,BallonBordo:2020jtw,Amirabi:2020mzv,Bandos:2020hgy,Bokulic:2021dtz,Zhang:2021qga,Nomura:2021efi,Barrientos:2022bzm,Colipi-Marchant:2023awk,Ayon-Beato:2024vph,Barrientos:2024umq,Correa:2024qej,Bokulic:2025usc,Hale:2025veb}. Recently, it has been demonstrated that, in asymptotically AdS spacetimes with axionic fields, the ModMax parameter modifies the charge transport, in particular, the Hall angle and Nerst signal~\cite{Barrientos:2025rde}. In the context of Conformal Gravity, ModMax theory is a natural candidate to study self-gravitating nonlinear effects without breaking the symmetries of the gravity sector, similar to what happens with conformally-coupled scalar fields.

Another motivation for studying conformally invariant theories is the unification of gravity and electromagnetism through Weyl geometry~\cite{Weyl1,Weyl2}. The latter appears as a generalization of Riemannian geometry by introducing a covariant derivative with a Weyl connection. Dirac proposed a generalization of Weyl's theory, by introducing a real scalar field~\cite{Dirac:1973gk,Dirac:1975vq}. Cosmological applications of the latter were considered in detail in Refs.~\cite{Rosen:1982nr,Israelit:2010jc}, and further generalizations of the Weyl theory were considered in~\cite{Utiyama:1973nq,Utiyama:1974qu,Nishioka:1985wc}. In this context, static and spherically symmetric black hole solutions have been investigated in detail in Refs.~\cite{Panpanich:2018cxo,Yang:2022icz} using both numerical and analytical methods, while near-horizon instabilities in Weyl black holes were studied in~\cite{Kouniatalis:2025pxs}.

In this work, we study nonconformally Ricci-flat gravitational instantons in Conformal Gravity, both in vacuum and coupled to nonlinear fields that remain invariant under Weyl rescalings in four dimensions. In the former case, we compute the Noether-Wald charges of the Lorentzian extension of the Kerr-NUT-AdS metric with relaxed AdS boundary conditions in Conformal Gravity found in Ref.~\cite{Liu:2012xn}. After performing an analytic continuation into the Euclidean section, we obtain a curve in parametric space where the Weyl tensor becomes (anti)self-dual. As the Conformal Gravity action is a quadratic form of the Weyl tensor, this implies that the instanton saturates a Bogomol'nyi-Prasad-Sommerfield (BPS) bound that we obtain explicitly. Then, we consider a matter sector consisting of two different types of nonlinear fields: conformally coupled scalar fields with a self-interacting potential and ModMax electrodynamics. We obtain the generalization of the Taub-NUT-AdS and Eguchi-Hanson metrics found in Ref.~\cite{Corral:2021xsu}, and we show that the vanishing NUT charge limit leads to a generalized Riegert metric in the presence of conformal nonlinear matter fields. We study their global properties in detail, including the partition function, topological terms, and conserved quantities. 

This work is organized as follows. In Sec.~\ref{sec:conformal gravity}, we briefly review Conformal Gravity in four dimensions. Then, in Sec.~\ref{sec:Kerr-NUT-AdS}, we revisit the Lorentzian Kerr-NUT-AdS extension in Conformal Gravity and compute its conserved charges. In Sec.~\ref{sec:KNAdS-Euclidean}, we study the Euclidean continuation of this geometry, determine the curve in parameter space along which the solution becomes (anti)-self-dual, and compute the corresponding on-shell Euclidean action. In Sec.~\ref{sec:Obstruction conformally Einstein spaces}, by using the Dunajski-Tod theorem, which provides the necessary and sufficient conditions for (anti)self-dual spaces not to be conformally related to any Ricci-flat space, we show that the Kerr-NUT-AdS extension satisfies the conditions of this theorem. In Sec.~\ref{sec:coupling nonlinear matter}, we minimally couple nonlinear conformal matter to Conformal Gravity, namely conformally coupled scalar fields and ModMax nonlinear electrodynamics. Section~\ref{sec:new gravitational instantons} is devoted to the analysis of the extensions of the Taub-NUT and Eguchi-Hanson gravitational instantons within this theory. Furthermore, by taking the static limit of the former, we obtain the generalization of the Riegert metric supported by nonlinear conformal matter. We analyze the global properties of these new solutions, including their on-shell Euclidean action, temperature, topological invariants, and conserved charges derived from a Gauss law. Finally, in Sec.~\ref{sec:discussion} we present a summary and discussion of the main results.

\section{Conformal gravity}\label{sec:conformal gravity}

In this section, we review the main aspects of Conformal Gravity and fix notation. The action principle that dictates the dynamics of the gravitational sector is given by
\begin{align}\label{ICG}
    I_{\rm CG}[g_{\mu\nu}] = \alpha\,\int_{\mathcal{M}}\diff{^4x}\sqrt{|g|}\;W^{\mu\nu}_{\lambda\rho} \, W_{\mu\nu}^{\lambda\rho}\,,
\end{align}
where $\alpha$ is a dimensionless coupling constant, $g=\det g_{\mu\nu}$ is the metric determinant, and the Weyl tensor is defined as
\begin{align}\label{Weyl-Schouten}
    W^{\mu\nu}_{\lambda\rho} = R^{\mu\nu}_{\lambda\rho} + 4\delta^{[\mu}_{[\lambda}S^{\nu]}_{\rho]}\,, \;\;\;\;\; \mbox{where} \;\;\;\;\; S^\mu_\nu = \frac{1}{2}\left(R^\mu_\nu - \frac{1}{6}\delta^\mu_\nu R \right)\,,
\end{align}
is the Schouten tensor. Besides diffeomorphism invariance, this action remains invariant under Weyl rescalings of the metric, since the Weyl tensor transforms as $W^{\mu\nu}_{\lambda\rho}\to\bar{W}^{\mu\nu}_{\lambda\rho}=e^{-2\sigma(x)}W^{\mu\nu}_{\lambda\rho}$ under
\begin{align}\label{WTmetric}
    g_{\mu\nu} \to \bar{g}_{\mu\nu} = e^{2\sigma(x)}g_{\mu\nu}\,,
\end{align}
where $\sigma(x)$ is an arbitrary scalar function. Although this action produces four-order field equations, it has been shown that Einstein gravity can be obtained by assuming Neumann boundary conditions on the Fefferman-Graham expansion~\cite{Maldacena:2011mk,Anastasiou:2016jix}. Moreover, the latter choice eliminates possible ghost instabilities as shown in Ref.~\cite{Hell:2023rbf}. 

The field equations are obtained by performing arbitrary variations of the action~\eqref{ICG} with respect to the metric, giving
\begin{align}
    \delta I_{\rm CG} = \int_{\mathcal{M}}\dd^4x\sqrt{|g|}\;\delta g^{\mu\nu}\mathcal{E}_{\mu\nu} + \int_{\mathcal{M}}\dd^4x\sqrt{|g|}\;\nabla_\mu \Theta^\mu\,,
\end{align}
where $\mathcal{E}_{\mu\nu}=0$ and $\Theta^\mu$ are the field equations and symplectic potential, given by
\begin{align}\label{eomvacuum}
    \mathcal{E}_{\mu\nu} := 4\alpha B_{\mu\nu} \,, \;\;\;\;\; \mbox{and} \;\;\;\;\; 
    \Theta^\mu := -4\alpha\left( \delta g_{\nu \sigma} C^{\sigma \mu \nu} +  W^{\rho \sigma \mu \nu} \nabla_\rho \delta g_{\nu \sigma}\right)\,,
\end{align}
respectively, where $B_{\mu\nu}=\nabla^\lambda C_{\mu\nu\lambda}-S^{\lambda\rho}W_{\lambda\mu\nu \rho}$ is the Bach tensor and $C_{\mu\nu\lambda}=2\nabla_{[\lambda}S_{\nu]\mu}$ is the Cotton tensor. Notice that, when evaluated on Einstein-AdS spaces, say $R_{\mu\nu}=-\tfrac{3}{\ell^2}g_{\mu\nu}$ the Schouten tensor becomes $S_{\mu\nu}=-\tfrac{1}{2\ell^2}g_{\mu\nu}$, where $\ell$ is the AdS radius. One can check that the Bach and Cotton tensors vanish automatically when this condition is imposed. Therefore, all Einstein-AdS spaces are Bach-flat, and as a consequence, solutions of Conformal Gravity in vacuum. Even more: as the Bach tensor transforms covariantly under Weyl transformations, conformally-Einstein spaces will also satisfy $B_{\mu\nu}=0$ for an arbitrary conformal factor. Nevertheless, there exists a very special class of metrics, which are Bach flat, but they are not conformally related to Einstein spaces. Necessary and sufficient conditions for their existence have been explored~\cite{Kozameh:1985jv,Gover:2004ab,Dunajski:2013zta} (see also~\cite{Lu:2013hx}), and particular examples, e.g. the LeBrun metrics, have been obtained~\cite{Lebrun:1991,LeBrun:1994}. 
 
In the following, we will analyze the electric/magnetic duality of a Bach-flat solution that can be regarded as the one-parameter extension of the Kerr-NUT-AdS instanton in Einstein gravity.  

\section{Kerr-NUT-AdS solution in Conformal Gravity}\label{sec:Kerr-NUT-AdS}

The construction of stationary solutions in higher-curvature theories of gravity is certainly challenging. Since Conformal Gravity has more symmetries than General Relativity, but it reduces to the latter by imposing Neumann boundary conditions on the Fefferman-Graham expansion, it is more likely to find this class of solutions in both Lorentzian and Euclidean signatures. For instance, in Ref.~\cite{Mannheim:1990ya}, the extension of the dyonic Kerr-Newman-AdS solution in Conformal Gravity was studied by using the Plebanski-Demianski form.\footnote{See Ref.~\cite{Asuncion:2025cxw} for a recent analysis in the absence of the cosmological constant and Maxwell fields.} Then, in Ref.~\cite{Liu:2012xn}, the dyonic Kerr-Newman-NUT-AdS was studied. Using Boyer-Lindquist coordinates, the ansatz for the line element is
    \begin{align}\notag
    \diff{s^2} &= - \frac{\Delta_r(r)}{\Xi^2\Sigma(r,\vartheta)}\left[\diff{t} + \left(2n\cos\vartheta+2nC-a\sin^2\vartheta\right)\diff{\varphi} \right]^2 + \frac{\Sigma(r,\vartheta)}{\Delta_r(r)}\diff{r^2} \\
    &\quad + \frac{\Sigma(r,\vartheta)}{\Delta_\vartheta(\vartheta)}\diff{\vartheta^2} + \frac{\Delta_\vartheta(\vartheta)\sin^2\vartheta}{\Xi^2\Sigma(r,\vartheta)}\left[a\diff{t}-\left(r^2+a^2+n^2-2anC \right)\diff{\varphi} \right]^2\,, \label{ansatz}
\end{align}
where $\Xi=1-a^2/\ell^2$, with $a$ being the rotation parameter, $n$ is the NUT charge, and $C=\pm1,0$ controls the position of the Misner string---the gravitational analog of the Dirac string. The $C$ parameter appears from a large gauge transformation that modifies the conserved charges, as we will see next. Indeed, as shown by Misner, it can be used to define two regular coordinate patches such that the string is fully eliminated, paying the price of introducing closed timelike curves~\cite{Misner:1963fr}. Here, however, we keep the Misner string physical, as it has been shown that it is transparent to geodesic observers~\cite{Clement:2015cxa}. The latter has provided a fruitful arena for studying novel thermodynamic properties in the presence of the NUT charge~\cite{Hennigar:2019ive,Bordo:2019slw,Bordo:2019tyh,BallonBordo:2019vrn,Durka:2019ajz,Ciambelli:2020qny}.   

The metric functions that solve the field equations~\eqref{eomvacuum} are
\begin{subequations}\label{solutionsvacuum}
\begin{align}\notag
\Delta_r(r) &= (c+1) \, r^2 - 2 \left( m + \frac{c \left( a^2 - 4 \, n^2 \right)}{3 \, \ell^2 \, b} \right) r + \frac{(a^2-n^2)(c+3)}{3} \\ &+ \frac{r^4 + (a^2+6 \, n^2) \, r^2 + 3 \, n^2 \, (a^2-n^2)}{\ell^2} + b \, r^3 \,,  \\
\Sigma(r,\vartheta) &= r^2 + \left(n+a\cos\vartheta \right)^2 \,, \\
\Delta_\vartheta(\vartheta) &= 1 - \left( \frac{ac}{3n} + \frac{4an}{\ell^2} \right) \cos\vartheta - \frac{a^2}{\ell^2} \cos\vartheta^2 \,,
\end{align}
\end{subequations}
where $m$, $b$, and $c$ are integration constants subject to the following constraint
\begin{align}\label{bsol}
   18mn^2 b -c\left[c \left( a^2 - 3n^2 \right) - 6n^2 \right] =0\,.
\end{align}

There are interesting limits of this metric. For instance, the Kerr-NUT-AdS solution of general relativity is also a solution of Conformal Gravity. The latter can be achieved by taking the limit $c\to0$, which implies that $b\to0$ as a consequence of Eq.~\eqref{bsol}. Notice that Eq.~\eqref{bsol} implies that the limit $c/b\to0$ is smooth. In Einstein gravity, this spacetime was originally found by Carter in Ref.~\cite{Carter:1968ks}, whose global properties were studied by Miller in~\cite{Miller:1973hqu}. Here, we notice that the limit $m\to0$ and $c\to0$, while keeping $b\neq0$, leads to a non-Einstein spacetime with nontrivial mass and angular momentum---see Eq.~\eqref{Q-KNNAdS} below. The general solution in Eq.~\eqref{solutionsvacuum} with nonvanishing $m$, $b$, and $c$, alongside their particular limits, possesses two Killing vectors, i.e., $\partial_t$ and $\partial_\varphi$, being stationary and axisymmetric. Additionally, it has a horizon generated by the Killing vector field $\xi=\partial_t + \Omega_+\partial_\varphi$, where
\begin{align}
    \Omega_+ = \frac{a}{r_+^2+a^2+n^2-2anC}\,,
\end{align}
is the angular velocity at the horizon, and $r_+$ is defined as the largest positive root of $\Delta_r(r_+)=0$. The Hawking temperature, on the other hand, can be obtained from $T_H = \frac{\kappa_s}{2\pi}$, where $\kappa_s$ is the surface gravity associated with the Killing vector field that generates the horizon. Direct computation of the latter yields
\begin{align}\notag
    T_H &= \frac{\Omega_+}{12\pi r_+ a\ell^2\Xi}\bigg\{\left(a^2-n^2\right)\left[(c+3)\ell^2+9n^2\right]\\ &\qquad \qquad\qquad-3r_+^2\left[(c+1)\ell^2+a^2+6n^2+r_+\left(3r_+ + 2b\ell^2\right)\right] \bigg\}
\end{align}
The thermodynamic properties of this solution have been studied in Ref.~\cite{Liu:2012xn} in the absence of the NUT charge. The presence of the latter modifies both the mass and the angular momentum, as we see next. Even more, in the Euclidean section, it allows for a nontrivial curve in parameter space where the solution becomes (anti)self-dual. This allows us to interpret it as a gravitational instanton of Conformal Gravity that saturates a BPS bound given in terms of the Chern-Pontryagin index.

\subsection{Noether-Wald charges}

As Conformal Gravity is finite in asymptotically locally AdS spacetimes~\cite{Grumiller:2013mxa}, the Noether-Wald formalism~\cite{Lee:1990nz,Wald:1993nt,Iyer:1994ys,Wald:1999wa} provides a useful covariant tool for obtaining the conserved charges of the solution~\eqref{solutionsvacuum}. This method is based on the invariance under diffeomorphisms of the Conformal Gravity action by the flow of a Killing vector field $\xi=\xi^\mu\partial_\mu$, which implies the conservation of a Noether current, say $\nabla_\mu J^\mu = 0$. The Poincaré lemma, in turn, allows one to write the Noether current as $J^\mu = \nabla_\nu q^{\mu\nu}$, where $q^{\mu\nu}$ is known as the Noether prepotential. In the case of Conformal Gravity, this is given by~\cite{Corral:2021xsu}
\begin{align}
    q^{\mu\nu} = -4\alpha (W^{\mu \nu}_{\lambda \rho} \nabla^\lambda \xi^\rho + 2 \xi^\lambda C_{\lambda}{}^{\mu\nu}) = -q^{\nu\mu}\,.
\end{align}
The non-Einstein contribution to the conserved charges is encoded in the last term proportional to the Cotton tensor. Indeed, for Einstein spaces, the latter vanishes, and the Noether prepotential reduces to the topologically renormalized Einstein-AdS gravity~\cite{Aros:1999id,Olea:2006vd,Miskovic:2009bm}, also known as \emph{Kounterterms}. If $\xi$ is a globally defined Killing vector field, conserved charges are obtained by integrating the Noether prepotential over a codimension-2 boundary integral as
\begin{align}
    \mathcal{Q}[\xi] = \int_{\Sigma_\infty}\,q^{\mu\nu}\diff{\Sigma_{\mu\nu}}\,,
\end{align}
where $\Sigma_\infty$ is the codimension-2 asymptotic boundary, whose oriented volume element is $\diff{\Sigma_{\mu\nu}}$. 

The mass and angular momentum of the solution in Eq.~\eqref{solutionsvacuum} are obtained from the boundary integrals at spacelike infinity associated with the Killing vectors $\partial_t$ and $\partial_\varphi$, giving
\begin{subequations}\label{Q-KNNAdS}
\begin{align}
    \mathcal{Q}[\partial_t] &:= M = \frac{64\pi\alpha}{\Xi^2\ell^2}\left[ m \, - \, \frac{n \left( C \, a + n \right) b}{2} \, + \frac{c}{12 \, b} \left( b^2 \ell^2 + \frac{4 \left( a^2 - 4 \, n^2 \right)}{\ell^2} \right) \right]\,, \\
    \mathcal{Q}[\partial_\varphi] &:= J = M(3Cn-a) + \frac{64\pi\alpha b}{\Xi^2\ell^2}\left[  \frac{n^2 \left(C^2 a + 2Cn - 2a\right)}{2} + \ell^2 \left( \frac{(c - 6) a}{12} - \frac{Ccn}{12} \right)\right]\,,
\end{align}    
\end{subequations}
respectively. This result coincides exactly with that found in Ref.~\cite{Corral:2024lva} in the limit $b\to0$ and $c\to0$, provided that $\alpha=\tfrac{\ell^2}{64\pi}$. Remarkably, notice that due to the conformal-gravity modes, there exists a solution with nontrivial mass in the $m\to0$ limit, due to the presence of $b$ and $c$. Additionally, the parameter associated with the large gauge transformation that controls the position of the Misner string contributes to the conserved charges, as anticipated. This is reminiscent of the large coordinate transformation that alters its asymptotic behavior, similar to what occurs with the cylindrical black hole in general relativity~\cite{Lemos:1994xp}.

\section{(Anti)Self-dual instanton in Conformal Gravity}\label{sec:KNAdS-Euclidean}

Gravitational instantons in Conformal Gravity are regular Bach-flat solutions that can be obtained from their Lorentzian counterpart by performing the analytic continuation into the Euclidean section. In particular, from Eq.~\eqref{solutionsvacuum}, the (anti)self-dual gravitation instanton can be obtained from the Wick rotation $t\to i\tau$, $n\to -i\hat{n}$, and $a\to -i\hat{a}$, where hatted quantities denote those in Euclidean signature. The absence of conical singularities renders the solution regular; this is achieved by demanding the correct periodicity of the Euclidean time and the azimuthal angle, i.e. $\tau\sim\tau+\beta_\tau$ and $\varphi\sim\varphi+\beta_\varphi$, where $\beta_\varphi:=-\beta_\tau\hat{\Omega}_+$ and $\beta_\tau = \hat{T}_H^{-1}$, with $\hat{\Omega}_+$ and $\hat{T}_H$ being the analytically continued angular velocity and Hawking temperature, respectively.  

Since the Conformal Gravity action is a quadratic form of the Weyl tensor in four dimensions, it can be used to define a BPS bound similar to Yang-Mills theory. To the best of our knowledge, this was first explored in the context of Conformal Gravity in Ref.~\cite{Deser:1983wb}. Indeed, in Ref.~\cite{Strominger:1984zy} the explicit value of the BPS bound was obtained for $\mathbb{CP}^2$ instanton. In the case of the analytically continued solution in Eq.~\eqref{solutionsvacuum}, there exists a curve in parameter space that saturates a different BPS bound. The latter is characterized by 
\begin{align}\label{SDcondition}
    m = \pm \frac{\hat{n}(\hat{a}^2+\ell^2-4\hat{n}^2)}{\ell^2} +\frac{c}{3}\left[\frac{(\hat{a}^2-4\hat{n}^2)}{\hat{b}\ell^2}\pm \frac{(3\hat{n}^2-\hat{a}^2)}{2\hat{n}}\right] \;\;\;\;\; \mbox{and}  \;\;\;\;\;  c = \mp 3\hat{n}\hat{b}\,,
\end{align}
where $\hat{b}$ is obtained by taking the Euclidean continuation of Eq.~\eqref{bsol}. The condition~\eqref{SDcondition} implies that the Weyl tensor is (anti-)self dual, i.e. $W_{\mu\nu\lambda\rho}=\pm\tilde{W}_{\mu\nu\lambda\rho}$, where $\tilde{W}_{\mu\nu\lambda\rho}:=\tfrac{1}{2}\varepsilon_{\mu\nu\alpha\beta}W^{\alpha\beta}_{\lambda\rho}$ is the dual Weyl tensor and $\epsilon_{\mu\nu\lambda\rho}$ the Levi-Civita tensor. For (anti)self-dual configurations, the BPS bound in Conformal Gravity is given by~\cite{Deser:1983wb,Strominger:1984zy}
\begin{align}\label{BPSbound}
    I_{\rm CG} = \pm 16\pi^2\alpha\,P_1[\mathcal{M}] \,,  \;\;\;\;\; \mbox{where} \;\;\;\;\;  P_1[\mathcal{M}] = \frac{1}{16\pi^2}\int_{\mathcal{M}}\dd^4x\sqrt{|g|}\;\tilde{R}_{\mu\nu\lambda\rho}R^{\mu\nu\lambda\rho}\,,
\end{align}
is the Chern-Pontryagin index, and $\tilde{R}_{\mu\nu\lambda\rho}=\tfrac{1}{2}\varepsilon_{\mu\nu\alpha\beta}R^{\alpha\beta}_{\lambda\rho}$ is the dual Riemann tensor. Indeed, the action of this instanton can be lower than that of $\mathbb{CP}^2$ within a particular range along the curve~\eqref{SDcondition}, meaning that the solution studied here dominates the path integral. As the Chern-Pontryagin index of our solution is
\begin{align}
P_1[\mathcal{M}]&=\frac{\beta_\tau\beta_\varphi}{\Xi^2\pi^2\ell^4[\hat{a}^2-(\hat{n}\pm r_+)^2]^3}\Big\{-2\hat{n}^2(\hat{a}^2+\ell^2-4\hat{n}^2)^2[\pm 2r_+(\hat{a}^2+r_+^2)+\hat{n}(\hat{a}^2+3r_+^2-\hat{n}^2)]\notag\\
	&\mp 2\hat{b}\hat{n}\ell^2(\hat{a}^2+\ell^2-4\hat{n}^2)[\hat{a}^2-(\hat{n}\mp r_+)^2][\hat{a}^2(\hat{n}\pm 2r_+)-3\hat{n}(\hat{n}\pm r_+)^2-\hat{b}\ell^2\hat{n}^2(\hat{a}-\hat{n}^2)(\hat{a}-2\hat{n}^2)]\notag\\
	&-\hat{b}^2\ell^4[r_+^5(\hat{a}^2+3\hat{n}^2)+2r_+^3[\hat{n}^2(\hat{n}^2+2\hat{a}^2)-\hat{a}^4]\pm 6r_+^2\hat{n}^3(2\hat{n}^2-\hat{a}^2)+r_+[(\hat{a}^2-\hat{n}^2)^3+4\hat{n}^6]\Big\}\,,
\end{align}
and that of the complex projective space is $P_1[\mathbb{CP}^2]=3$, we conclude that, since the parametric space of this (anti)-self dual instanton is two-dimensional, there is always a region in the parameter space such that the condition $P_1[\mathcal{M}]\leq P_1[\mathbb{CP}^2]$ is satisfied. Therefore, within such a region, the (anti)self-dual instanton studied here would dominate the path integral if the one-loop determinant is not imaginary. Indeed, one can use the techniques developed in Ref.~\cite{Strominger:1984zy} for gravitational instantons in Conformal Gravity to prove this. We left this proof for future work.

\section{Obstructions to conformally (anti)self-dual Ricci-flat spaces}\label{sec:Obstruction conformally Einstein spaces}

It is well-known that the conformal symmetry of the action~\eqref{ICG} allows one to obtain Bach-flat solutions by performing a Weyl transformation of Einstein spaces. This is because the Bach tensor is identically zero for Einstein spaces, and it transforms covariantly under Weyl rescalings of the metric [cf. Eq.~\eqref{WTmetric}] with conformal weight $\Delta=-2$, that is, $B_{\mu\nu}\to\bar{B}_{\mu\nu}=e^{-2\sigma(x)}B_{\mu\nu}$. Thus, if $\bar{g}_{\mu\nu}$ is conformally related to an Einstein space, it will solve the Conformal Gravity equations automatically. However, as demonstrated in Refs.~\cite{Kozameh:1985jv,Gover:2004ab,Dunajski:2013zta}, there exist solutions of Conformal Gravity which are not conformally Ricci-flat. Indeed, in Ref.~\cite{Dunajski:2013zta} it was shown that a Bach-flat Riemannian metric with anti-self-dual Weyl tensor contains a Ricci-flat metric within its conformal class if and only if
\begin{subequations}\label{constraints-confE}
\begin{align}
\mathcal{P}:=4 \nabla^\mu W_{\alpha\beta\gamma\mu} \nabla_\nu W^{\alpha\beta\gamma\nu} - V_\mu V^\mu W_{\alpha\beta\gamma\delta}W^{\alpha\beta\gamma\delta} &= 0\,, \label{no-confE-cons1}\\
\mathcal{T}_{\mu\nu}:=S_{\mu\nu} + \nabla_\mu V_\nu + V_\mu V_\nu - \frac{1}{2} g_{\mu\nu}V_\lambda V^\lambda &= 0,\label{no-confE-cons2}
\end{align}    
\end{subequations}
where $V^\mu=4W^{\sigma\mu}_{\lambda\rho}\nabla^\nu W^{\lambda\rho}_{\sigma\nu}/W^2$ and $W^2 = W_{\mu\nu\lambda\rho}W^{\mu\nu\lambda\rho}\neq0$. This is the Dunajski-Tod theorem~\cite{Dunajski:2013zta}.

In particular, for the one-parameter extension of the Kerr-NUT-AdS solution in Eqs.~\eqref{ansatz} and~\eqref{solutionsvacuum}, the scalar in Eq.~\eqref{no-confE-cons1} takes the form
\begin{equation}
\mathcal{P}=-\frac{18b^2}{\ell^2\,(n - r + a\,\cos\vartheta)^7}
\Bigg[
  \beta_1(r)\Big(n^2(n + r) + a n(n + 2r) \cos\vartheta\Big)
  + a^{2} \beta_2(r,\vartheta)
\Bigg],
\end{equation}
where we have defined
\begin{subequations}
\begin{align}
\beta_1(r) &= -\big(3n - r\big)\big(n + r\big) + \ell^{2}\big(1 + b(n + r)\big)\,,\\
\beta_2(r,\vartheta) &= (3n - r)(n^2 - r^2 \cos^2\vartheta)
    + \ell^2\big(-n(1 + bn) + r + b\,r^2\cos^2\vartheta\big)\notag\\&~~~
    - a\cos\vartheta\big(-3\,n^2 - 2nr + \ell^2\big[1 + b(n + r)\big] + r^2\cos^2\vartheta\big)\,.
 \end{align}
\end{subequations}
This expression is nonvanishing as long as $b\neq 0$, even in the limit $\ell\to\infty$. Therefore, the nonvanishing of Eq.~\eqref{constraints-confE} implies that the solution is not conformal to a Ricci-flat space. In fact, switching off the integration constant $b$ also forces $c=0$, according to Eq.~\eqref{SDcondition}, thereby recovering the Kerr-NUT-AdS solution of General Relativity. Additionally, when evaluated on the one-parameter extension of the latter, the tensor $\mathcal{T}_{\mu\nu}$ in Eq.~\eqref{no-confE-cons2} can be expressed as
\begin{equation}
	\mathcal{T}_{\mu\nu}=H(r,\vartheta)g_{\mu\nu}\,,
\end{equation}
where 
\begin{equation}
	H(r,\vartheta)=\pm \frac{(\pm \hat{b}\ell^2\hat{n}-\hat{a}^2-\ell^2+4\hat{n}^2)[\hat{b}^2\ell^4(\hat{a}^2-\hat{n}^2)\pm 4\hat{b}\ell^2\hat{n}(\hat{a}^2-2\hat{n}^2)+4\hat{n}^2(\hat{a}^2+\ell^2-4\hat{n}^2)]}{2\ell^2[\hat{b}\ell^2\hat{a}(\hat{a}\pm r\cos\vartheta)+\hat{n}(\hat{b}\ell^2r+2(\hat{a}^2+\ell^2))-\hat{n}^2(\pm3\hat{b}\ell^2+8\hat{n})]^2}\,,
\end{equation}
is a monotonically decreasing function of the radial coordinate, whose falloff toward the asymptotic boundary is $H\sim\mathcal{O}(r^{-2})$. For arbitrary values of the integration constants, this scalar is nonvanishing, even in the limit $\ell\to\infty$ and $b\neq0$. Therefore, we conclude that the Kerr-NUT-AdS instanton of Conformal Gravity is not conformally related to its Ricci-flat counterpart, as long as $b\neq0$. This result extends the nonconformally Einstein metrics of Conformal Gravity studied in Ref.~\cite{Liu:2013fna}. Even more, as the one-parameter extension of the Taub-NUT-AdS solution is continuously connected to that studied here in the limit $a\to0$, one can conclude from the Dunajski-Tod theorem~\cite{Dunajski:2013zta} that the solutions studied in Ref.~\cite{Corral:2021xsu} are nonconformally Ricci-flat as well.

\section{Coupling to nonlinear conformal matter}\label{sec:coupling nonlinear matter}

To study the backreaction of matter fields on gravitational instantons in Conformal Gravity, the dynamics of the matter content must respect the conformal invariance of the gravity sector. To this end, we consider a matter content composed of two distinct types of nonlinear fields: conformally coupled scalar fields with a self-interacting potential and ModMax electrodynamics.\footnote{In Einstein gravity, gravitational instantons with conformally-coupled scalar fields or ModMax fields have been studied in Refs.~\cite{deHaro:2006wy,deHaro:2006ymc,Bardoux:2013swa,BallonBordo:2020jtw,Flores-Alfonso:2020nnd,Barrientos:2022yoz}.} The former is described by the action principle
\begin{align}\label{Iphi}
     I_\phi[g_{\mu\nu},\phi] &= -\int_{\mathcal{M}}\diff{^4x}\sqrt{|g|}\left(\frac{1}{2}\nabla_\mu\phi\nabla^\mu\phi + \frac{1}{12}\phi^2R + \nu\phi^4 \right) \,,
\end{align}
where $\nu$ is a dimensionless parameter that controls the quartic potential of the scalar field. This action remains quasi-invariant, that is, it transforms as a boundary term, under Weyl rescalings of the metric [cf. Eq.~\eqref{WTmetric}] and the scalar field $\phi\to\bar{\phi}=e^{-\sigma(x)}\phi$, namely (see Ref.~\cite{Anastasiou:2022wjq})
\begin{equation}\label{deltasigmaIphi}
    \delta_{\sigma} I_{\phi} = -\frac{1}{2} \int \mathrm{d}^4 x \sqrt{|g|} \nabla_{\mu} \left( \phi^2 \nabla^{\mu} \sigma \right)\,,
\end{equation}
where we have used the infinitesimal Weyl transformation of the metric and the scalar field as $\delta_\sigma g_{\mu\nu}=2\sigma g_{\mu\nu}$ and $\delta_\sigma \phi = -\sigma\phi$, respectively.

To obtain finite action and conserved charges in asymptotically locally AdS spacetimes, this action needs to be supplemented with additional counterterms that do not modify the bulk dynamics, while rendering the action fully invariant under Weyl rescalings. This is the conformal completion prescription for scalar-tensor theories used in Ref.~\cite{Anastasiou:2022wjq}, which we follow from hereon. To this end, we consider the Weyl-invariant action
\begin{align}\label{Iphiren}
    I^{\rm (ren)}_{\phi} = I_\phi -  \frac{1}{96 \nu} \int \dd^4 x \sqrt{|g|} \left( E_4 + \nabla_\mu J^\mu \right)\,,
\end{align}
where $I_\phi$ is defined in Eq.~\eqref{Iphi}, and $E_4$ is the Gauss-Bonnet density defined as 
\begin{align}\label{E4}
E_4 = \frac{1}{4}\delta^{\mu_1\ldots\mu_4}_{\nu_1\ldots\nu_4}R^{\nu_1\nu_2}_{\mu_1\mu_2}R^{\nu_3\nu_4}_{\mu_3\mu_4} =  R^2 - 4R_{\mu\nu}R^{\mu\nu} + R_{\mu\nu\lambda\rho}R^{\mu\nu\lambda\rho}\,, 
\end{align}
where $\delta^{\mu_1\ldots\mu_p}_{\nu_1\ldots\nu_p}=p!\delta^{[\mu_1}_{[\nu_1}\dots\delta^{\mu_p]}_{\nu_p]}$ is the generalized Korenecker delta. Additionally, since neither the Gauss-Bonnet term nor the action~\eqref{Iphi} are invariant under Weyl rescalings, the last term
\begin{align}
    J^\mu &= 8 \left[ \phi^{-1} G_{\lambda}^{\mu}  + \phi^{-2} \delta^{\mu\nu}_{\lambda\rho}\nabla^\rho\nabla_\nu\phi + \phi^{-3} \nabla^\mu \phi \nabla_\lambda \phi \right]\nabla^{\lambda} \phi - 48\nu\phi \nabla^\mu \phi  := J^\mu_{\rm E}  - 48\nu \phi \nabla^\mu \phi \,,\label{Jmu}
\end{align}
is included to render the renormalized scalar-tensor sector conformally invariant. The last piece of $J^\mu$ compensates for the nontrivial Weyl variation of $I_\phi$---see Eq.~\eqref{deltasigmaIphi}. On the other hand, $J^\mu_{\rm E}$ compensates the inhomogeneous piece of the Gauss-Bonnet term. This was shown in Ref.~\cite{Oliva:2011np} by introducing the tensor 
\begin{equation}\label{SouryaTensor}
    S^{\mu \nu}_{\lambda \rho} = \phi^2 R^{\mu \nu}_{\lambda \rho} 
    - 4 \phi \delta^{[\mu}_{[\lambda} \nabla^{\nu]} \nabla_{\rho]} \phi 
    + 8 \delta^{[\mu}_{[\lambda} \nabla^{\nu]} \phi \nabla_{\rho]} \phi 
    - \delta^{\mu \nu}_{\lambda \rho} \nabla_\alpha \phi \nabla^\alpha \phi\,,
\end{equation}
which transforms covariantly under Weyl rescalings, i.e. $ \mathcal{S}^{\mu \nu}_{\lambda \rho}\to\bar{\mathcal{S}}^{\mu\nu}_{\lambda\rho} = e^{-4\sigma(x)} $. Then, the product of $\phi^{-4}$ and the Euler density constructed out of Eq.~\eqref{SouryaTensor}, namely,
\begin{align}
   E_\mathcal{S} := \frac{1}{4\phi^4}\delta^{\mu_1\ldots\mu_4}_{\nu_1\ldots\nu_4}\mathcal{S}^{\nu_1\nu_2}_{\mu_1\mu_2}\mathcal{S}^{\nu_3\nu_4}_{\mu_3\mu_4} = E_4 + \nabla_\mu J^\mu_{\rm E}\,,
\end{align}
is conformally covariant by construction; in particular, it transforms as $E_{\mathcal{S}} \to \bar{E}_{\mathcal{S}}=e^{-4\sigma(x)}E_{\mathcal{S}}$. Then, since $E_4$ does not transform covariantly but $E_{\mathcal{S}}$ does, it is clear that $\nabla_\mu J^\mu_{\rm E}$ compensates the inhomogeneous Weyl transformation of the Gauss-Bonnet term.

On the other hand, in order to introduce nonlinear Abelian effects while respecting conformal invariance, we consider ModMax electrodynamics, whose action principle is given by~\cite{Bandos:2020jsw} 
\begin{align}\label{IMM}
    I_{\rm MM}[g_{\mu\nu},A] &= -\int_{\mathcal{M}}\diff{^4x}\sqrt{|g|}\left(X\cosh\gamma - \sqrt{X^2 - Y^2}\,\sinh\gamma \right)\,,
\end{align}
where $\gamma$ is a dimensionless parameter that controls the strength of the nonlinear effects. Indeed, Maxwell electrodynamics is continuously connected to ModMax theory in the limit $\gamma\to0$. Here, $X$ and $Y$ are the two quadratic electromagnetic invariants constructed out of the Abelian field strength $F_{\mu\nu}=\partial_\mu A_\nu - \partial_\nu A_\mu$ and its dual $\tilde{F}_{\mu\nu}=\frac{1}{2}\epsilon_{\mu\nu\alpha\beta}F^{\alpha\beta}$, where $\epsilon_{\mu\nu\alpha\beta}$ is the Levi-Civita tensor. Explicitly, they are given by
\begin{equation}
    X=\frac{1}{4}F_{\mu\nu}F^{\mu\nu}\qquad \text{and}\qquad Y=\frac{1}{4}\tilde{F}_{\mu\nu}F^{\mu\nu}\,.
\end{equation}
The action~\eqref{IMM} describes the most general nonlinear electrodynamic theory that remains invariant under $SO(2)$ duality transformations and Weyl rescalings~\cite{Bandos:2020jsw}. 

Then, the full action that governs the dynamics of Conformal Gravity coupled to nonlinear conformal matter fields is given by
\begin{align}\label{Itotal}
     I = I_{\rm CG} + I_\phi^{\rm (ren)} + I_{\rm MM}\,,
\end{align}
where the definitions of Eqs.~\eqref{ICG},~\eqref{Iphiren}, and~\eqref{IMM} have been used. The field equations of the system are obtained by performing arbitrary variations with respect to the metric, scalar, and nonlinear Abelian field; they are
\begin{subequations}\label{eom}
\begin{align}\label{eom-g}
    \mathcal{E}_{\mu\nu} &:= 8\alpha B_{\mu\nu} - T^{\rm (A)}_{\mu\nu} - T^{(\phi)}_{\mu\nu} = 0\,,  \\
    \label{eom-KG}
    \mathcal{E} &:= \Box\phi-\frac{1}{6}R\phi-4\nu\phi^3=0\,, \\ 
    \label{eom-MM}
    \mathcal{E}^\nu &:= \nabla_\mu P^{\mu\nu} =0\,, 
\end{align}    
\end{subequations}
respectively, where we have defined the constitutive tensor and stress-energy momentum tensor for scalar and ModMax fields, respectively, as
\begin{align}\label{Pmunu}
    P_{\mu\nu}&=\left(\cosh\gamma-\frac{X\sinh\gamma}{\sqrt{X^2-Y^2}}\right)F_{\mu\nu}+\frac{Y\sinh\gamma}{\sqrt{X^2-Y^2}}\tilde{F}_{\mu\nu}\,, \\
    T_{\mu\nu }^{\rm (A)}&=P_{(\mu}^{~\lambda}F_{\nu)\lambda}-g_{\mu\nu }\left(X\cosh\gamma-\sqrt{X^2-Y^2}\sinh\gamma\right)\,,\\
  T_{\mu\nu}^{(\phi)}&=\nabla_\mu\phi\nabla_\nu\phi -\frac{1}{2}g_{\mu\nu }\nabla_\alpha\phi\nabla^\alpha\phi +\frac{1}{6}(g_{\mu\nu }\Box-\nabla_\mu\nabla_\nu +G_{\mu\nu })\phi^2-\nu g_{\mu\nu}\phi^4\,.
\end{align}

As discussed in Ref.~\cite{Anastasiou:2022wjq}, the Gauss-Bonnet term is needed to renormalize the scalar-tensor sector when the scalar field is $\phi=\phi_0=$ constant. Notice that this last condition implies that the field equation for the scalar field~\eqref{eom-KG} becomes $R=-12/\ell_0^2$, where the effective AdS radius is defined in terms of the constant scalar field as $\ell_0^{-2}:=2\nu\phi_0^2$. The constant scalar field, however, breaks the conformal invariance. This can be observed by noting that the scalar-tensor sector becomes Einstein-AdS gravity, where the constant scalar field plays the role of an effective AdS radius. Indeed, neglecting ModMax terms, the field equations for the metric become
\begin{align}
G_{\mu\nu} - \frac{3}{\ell_0^2}g_{\mu\nu} - 48\alpha\ell_0^2 B_{\mu\nu} = 0\,,    
\end{align}
which are the field equations of Einstein-Weyl gravity. Indeed, there is a particular point in the parameter space, i.e. $\alpha=1/48$, where the theory reduces to Critical Gravity~\cite{Lu:2011zk}. At that point, the two spin-2 degrees of freedom becomes massless~\cite{Deser:2011xc}, resembling the chiral points of Topologically Massive Gravity~\cite{Deser:1982vy,Carlip:2008eq,Li:2008dq} and New Massive Gravity~\cite{Bergshoeff:2009hq,Bergshoeff:2009aq,Giribet:2010ed}. At that point, the conserved charges of Einstein spaces vanish, and they are nontrivial otherwise~\cite{Deser:2011xc,Miskovic:2014zja,Anastasiou:2017rjf}. 

\section{Nonlinearly charged Gravitational instantons}\label{sec:new gravitational instantons}

Besides constant scalar fields, we are interested in gravitational instantons with self-gravitating conformally coupled scalars and ModMax fields, which introduce nonlinear effects while preserving conformal symmetry. Recently, the extension of the Riegert black hole with meronic fields was found in Conformal Gravity minimally coupled to Yang-Mills theory~\cite{Flores-Alfonso:2024gag}. Here, we construct a novel gravitational Taub-NUT instanton with self-gravitating, nonlinear, and conformal matter. The solution can be regarded as a nonlinearly charged gravitational dyon, which is a gravitational solution with electric and magnetic mass. It represents a two-parameter extension of the original Taub-NUT in general relativity with a negative cosmological constant. Additionally, we also obtain a two-parameter extension of the Eguchi-Hanson space with a negative cosmological constant by introducing nonlinear matter in Conformal Gravity.

\subsection{Taub-NUT instanton with weakened AdS boundary conditions}\label{sec:Taub-NUT}

To solve the field equations and construct a nonlinearly charged two-parameter extension of the Taub-NUT-AdS instanton, we consider a metric ansatz based on the Hopf fibration over $\mathbb{S}^2$. In particular, we focus on the line element
\begin{align}
    \diff{s^2} = f(r)\left(\diff{\tau} + 2n\cos\vartheta\diff{\varphi} \right)^2 + \frac{\diff{r^2}}{f(r)} + (r^2-n^2)(\diff{\vartheta} + \sin^2\vartheta\diff{\varphi}^2)\,,
\end{align}
where $n$ is the NUT charge. Additionally, for the ModMax field, we assume an ansatz aligned along the Hopf fibration, that is,
\begin{align}
A=a(r)(\dd\tau+2n\cos\vartheta\dd\varphi)\,.    
\end{align}
Inserting these into the field equations, we find that they are solved by
\begin{subequations}\label{TN-solution}
\begin{align}
    f(r) &= \frac{r^2+n^2}{r^2-n^2} - \frac{2mr}{r^2-n^2} + \frac{r^4-6n^2r^2-3n^4}{\ell^2(r^2-n^2)} + \frac{c\left(r^2+\frac{n^2}{3} \right)}{r^2-n^2} + \frac{br^3}{r^2-n^2} \,, \\
    a(r) &= \frac{p}{2n}\cosh\left[e^\gamma\ln\left(\frac{r-n}{r+n} \right) \right] + \frac{q}{2n}\sinh\left[e^\gamma\ln\left(\frac{r-n}{r+n} \right) \right] \,, \\
    \phi(r) &= \frac{1}{r+n} \sqrt{\frac{c-3bn}{12\nu}} \,,
\end{align}    
\end{subequations}
where $m$, $\ell$, $b$, $c$, $q$, and $p$ are integration constants subject to the constraints
\begin{subequations}\label{constraints-TN}
\begin{align}
 0 &= 4n\ell^2c  -3\ell^2\left(b n^2 - 2m - 2n\right) - 24 n^3 \label{cons1-TN}\,, \\
 0 &= \frac{384 \nu n^2  (p^2-q^2) e^{\gamma}}{1+192\alpha}\nu +\left( bn^2 + 6m  - 2n + \frac{8 n^3}{\ell^2} \right)  \left(3 b n^2 + 2m   + 2 n - \frac{8 n^3}{\ell^2} \right) \label{const2-TN}\,.
\end{align}    
\end{subequations}
This metric is asymptotically locally AdS, as can be seen by analyzing the asymptotic behavior of the Riemann tensor toward the asymptotic boundary. Nevertheless, it has a weakened AdS asymptotics due to the presence of the $b$-mode.

The vacuum Taub-NUT-AdS solution of Conformal Gravity was studied in extenso in Ref.~\cite{Corral:2021xsu}. Here, the Weyl symmetry of the matter sector allows for a natural extension of the latter. Indeed, notice that the self-gravitating conformally-coupled scalar field is regular $\forall r\in\mathbb{R}_{\geq0}$ if $n>0$ and it depends critically on the presence of a quartic potential. Thus, nonlinear conformal effects are crucial for the existence of this solution. Moreover, the profile of the gauge potential over the Taub-NUT metric of Einstein-ModMax theory studied in Refs.~\cite{BallonBordo:2020jtw,Flores-Alfonso:2020nnd,Colipi-Marchant:2023awk} is similar to that found here. 

The absence of conical singularities at the horizon defined by the largest positive root of the polynomial $f(r_+)=0$ demands that the Euclidean time must be identified as $\tau\sim\tau+\beta$, whose period is given by
\begin{equation}\label{betanut}
    \beta = \eval{\frac{4\pi}{f'(r)}}_{r=r_+}
    = \frac{4\pi\ell^2 r_+(n^2-r_+^2)}{\ell^2[ c(n^2+r_+^2)-4Gmr_++3n^2+r_+^2]-(3n^2+r_+^2)^2}\,.
\end{equation}
This relation defines the reciprocal of the Hawking temperature of the nonlinearly charged Taub-NUT-AdS solution of Conformal Gravity.

Additionally, the Taub-NUT solution in Eq.~\eqref{TN-solution} exhibits a curve in parameter space along which it becomes globally (anti)-self-dual. This curve is characterized by two relations among the integration constants; they are
\begin{align}\label{sd-TN}
	m&=\pm \frac{n}{G}\left(1 -\frac{4n^2}{\ell^2} + \frac{c}{2}\right)\,, &  c&=\mp3bn\,, & p&=\pm q\,.
\end{align}
In the self-dual case, these conditions are translated into a relation among $b$, $n$, and $\ell$
in order to satisfy the constraints in Eq.~\eqref{const2-TN}. In the case $p=q=0$, this curve reduces to that obtained in Ref.~\cite{Corral:2021xsu}. Thus, Eq.~\eqref{sd-TN} represents its extension in the presence of dyonic charges. Notice that, along the self-dual curve, the scalar field vanishes identically for the anti-self-dual, and the ModMax field becomes self-dual in a nonlinear sense, i.e. $P_{\mu\nu}=\tilde{F}_{\mu\nu}$, where $P_{\mu\nu}$ is defined in Eq.~\eqref{Pmunu}. Indeed, one can check that the ModMax stress-energy tensor $T_{\mu\nu}^{(A)}$ vanishes, although the self-dual ModMax fields are nontrivial. This is typically what happens with instantons in Yang-Mills theory. Additionally, along the self-dual curve, the metric function takes the form~\cite{Corral:2021xsu}
\begin{equation}\label{TN-sd}
	f_{\rm NUT}(r)=\frac{r-n}{r+n}+\frac{b(r-n)^2}{r+n}+\frac{(3n+r)(r-n)^2}{(r+n)\ell^2}\,.
\end{equation}
Notice that, in this case, the metric possesses a zero-dimensional set of fixed points at $r=n$ where the metric degenerates --- this is usually referred to as a NUT. Additionally, it is completely regular in the range $n\leq r<\infty$ as long as the conditions~\eqref{sd-TN} are imposed on the period of the Euclidean time in Eq.~\eqref{betanut}. The latter gives $\beta_\tau=8\pi n$ and the Euler characteristic of this solution is $\chi(\mathcal{M})=1$. 

Notice that, in the absence of matter fields, this solution reduces to that studied in Ref.~\cite{Corral:2021xsu}. In such a case, it is worth mentioning that this self-dual instanton is not conformally Ricci-flat either, since the constraints in Eq.~\eqref{constraints-confE} reduce to
\begin{equation}
	\mathcal{T}_{\mu\nu}=\frac{[(bn-1)\ell^2+4n^2][(b\ell^2+4n^2)^2-4\ell^2]}{-2\ell^2[8n^2+3b\ell^2n-\ell^2(br+2)]^2}g_{\mu\nu}\,,
\end{equation}
while $\mathcal{P}=0$. The first condition is nonvanishing as long as $b\neq0$, and it remains valid even in the limit $\ell \to \infty$. In the case $b=0$, the constraint~\eqref{sd-TN} further implies $c=0$, and one recovers the standard self-dual Taub-NUT metric, which is an Einstein space. In other words, the presence of the parameters $b$ and $c$ prevents the solution from being conformally related to a Ricci-flat space. This analysis was not presented in Ref.~\cite{Corral:2021xsu}, and it provides a formal demonstration that the solution cannot be obtained from a conformal transformation of the Taub-NUT-AdS solution of Einstein gravity.

On the other hand, the Euclidean on-shell action for the self-dual Taub-NUT instanton sourced by nonlinear conformal matter, cf. Eq.~\eqref{TN-sd}, is
\begin{align}
	I_{\rm E}&=\frac{\pi^2}{72\nu }\left[(576q^2-3456\alpha)\nu+45+4c(c+3)-\frac{24n^2}{\ell^2}(2c-1152\nu\alpha+15)-\frac{144n^4}{\ell^4}(384\nu\alpha-5)\right],
\end{align}
where $\beta=8\pi n$ has been used. On the other hand, the Euclidean on-shell action for the Taub-Bolt solution can be obtained in a similar manner. However, its explicit form is cumbersome and not very illuminating. Nevertheless, it is worth mentioning that there appear logarithmic terms in the partition function, which are related to the presence of nonlinear terms of ModMax theory. A similar behavior has been observed in the Einstein-Skyrme model, where the nonlinear nature of Skyrmions can be used to construct deformations of $\mathbb{CP}^2$ as gravitational instantons in general relativity~\cite{Canfora:2025roy}.

In the case of ModMax theory, the field equations~\eqref{eom-MM} allow one to define the electric charge employing a Gauss law. Similarly, since the magnetic charge is a topological charge, it can be computed via the integral of the first Chern class. Then, these conserved charges are given by 
\begin{align}\label{QeQm}
    Q_e &= - \frac{1}{4\pi}\int_{\Sigma_\infty}\star P \;\;\;\;\; \mbox{and} \;\;\;\;\;  Q_m = - \frac{1}{4\pi}\int_{\Sigma_\infty} F\,,
\end{align}
where $\star$ is the Hodge dual, $\Sigma_\infty$ denotes spatial infinity, and we have defined $P=\tfrac{1}{2}P_{\mu\nu}\dd x^\mu\wedge\dd x^\nu$ and $F=\tfrac{1}{2}F_{\mu\nu}\dd x^\mu\wedge \dd x^\nu$ as the constitutive relation and $U(1)$ field strength $2$-forms, respectively. For ModMax fields over the extended Taub-NUT solution with weakened AdS boundary conditions, the electric and magnetic charges are given by
\begin{subequations}
	\begin{align}
	Q_e&= \lim_{r\to\infty}\left[(r^2-n^2)a'(r)e^{-\gamma}\right] =q\,,\\
	Q_m&=\lim_{r\to\infty}2na(r)=p\,,
\end{align}
\end{subequations}
respectively. Therefore, this solution represents a self-gravitating nonlinear dyon in Conformal Gravity that becomes self-dual along the curve~\eqref{sd-TN}.

\subsubsection*{Static limit: The conformally nonlinearly charged Riegert black hole}

The static limit can be obtained by taking $n\to 0$. At the level of the metric and the scalar field, this limit is trivial. For the constraints in Eq.~\eqref{constraints-TN}, however, this limit is subtle, as it requires a redefinition of the integration constants such that the limit is smooth. For instance, from Eq.~\eqref{cons1-TN}, we see that $m\to0$ as $n\to 0$. To take the limit smoothly, we redefine $m:=\mu n$, where $\mu$ is a rescaled integration constant. Then, substituting this into Eq.~\eqref{cons1-TN}, it gives
\begin{equation}
	4c\ell^2n+6\left(-\frac{1}{2}bn^2+\mu nG+n\right)\ell^2-24n^3=0\,.
\end{equation}
Then, to linear order in the $n\to0$ expansion, this condition fixes
$\mu$ in terms of the integration constant $c$ according to 
\begin{equation}\label{mmu}
\mu=-\frac{2c+3}{3G}\,.
\end{equation}
 Substituting this back into Eq.~\eqref{const2-TN} and expanding once again yields
\begin{equation}
	\left[384\ell^4\nu(p^2-q^2)e^\gamma+\frac{16}{3}\ell^2c(c+2)(192\alpha\nu+1)\right]n^2 + \mathcal{O}(n^4)=0\,.
\end{equation}
Then, to second order in $n$, and after some simplifications, we find that the restrictions on the integration constants are $m=0$ and
\begin{equation}\label{Riegertconstraint}
	(p^2-q^2)\nu e^{\gamma}+\frac{8c}{3}(c+2)\left(\nu\alpha+\frac{1}{192}\right)=0\,.
\end{equation}
Thus, if the restriction on the integration constants given in Eq.~\eqref{Riegertconstraint} is assumed, the solution reduces to the nonlinearly charged dyonic Riegert metric dressed by scalar fields, whose specific form is given by
\begin{subequations}\label{Reigertsolution}
\begin{align}
    \dd s^2 &= \left(\frac{r^2}{\ell^2} + br + 1 + c  \right) \dd \tau^2 + \frac{\dd r^2}{\left(\frac{r^2}{\ell^2} + br + 1 + c  \right)} + r^2\left(\dd\vartheta^2 + \sin^2\vartheta\dd\varphi^2 \right), \\ A &= \left(\Phi-\frac{q}{r}e^\gamma\right)\dd\tau +p\cos\vartheta\dd\varphi\,, \\ \phi&=\frac{1}{r}\sqrt{\frac{c}{12\nu}}\,,
\end{align}
\end{subequations}
where we kept the pure-gauge term $\Phi:=qe^\gamma/r_+$ to render the solution regular at the horizon. Notice that, from Eq.~\eqref{Riegertconstraint}, we can solve for $p=p(c,q)$, leaving two integration constants free, i.e. $c$ and $q$. They can be interpreted as a scalar hair and the electric charge, respectively. Indeed, the electric and magnetic charges can be obtained from the Gauss integrals in Eq.~\eqref{QeQm}, giving
\begin{equation}\label{ModMax-charges-Reigert}
	Q_e=q\,,\qquad Q_m=p\,.
\end{equation}

Some comments are now in order. This solution is the massless limit of the Riegert black hole~\cite{Riegert:1984zz} and it appears as a particular solution of Conformal Gravity in vacuum. Indeed, in Ref.~\cite{Lu:2012xu}, the authors showed that such a vacuum solution has a nontrivial Hawking temperature, but its partition function vanishes identically, resulting in a configuration with zero mass, entropy, and free energy; this is similar to what happens with the Schwarzschild-AdS black hole in critical gravity~\cite{Lu:2011zk} (see also~\cite{Anastasiou:2017rjf,Anastasiou:2021tlv}). In this case, the temperature can be obtained from Eq.~\eqref{betanut} by taking $m=\mu n$, with $\mu$ given by Eq.~\eqref{mmu}, and subsequently taking the limit $n\to 0$. This procedure yields the temperature of the static solution in Eq.~\eqref{Reigertsolution}, that is,
\begin{equation}
	\beta = \frac{4\pi\ell^2 r_+}{r_+^2-\ell^2(1+c)}\,.
\end{equation}
Therefore, due to its nonvanishing temperature in the absence of scalar and Maxwell fields, the solution was interpreted as a thermalized vacuum in Ref.~\cite{Lu:2012xu}.\footnote{A similar behavior was found in unconventional conformal supergravity~\cite{Alvarez:2022wcj}} Nevertheless, the presence of nonlinear conformal matter breaks the degeneracy and the partition function is no longer zero, as they contribute in a nontrivial way to the Euclidean on-shell action, which is given by
\begin{equation}
	-I_{\rm E}=\frac{\pi\beta}{36\nu r_+}\left[12+c(12+c+192c\nu\alpha )\right]-\frac{2\pi\beta e^\gamma(p^2+q^2)}{r_+}\,.
    \end{equation}
The first term on the right-hand side originates from the conformally coupled scalar sector, while the second corresponds to the contribution of ModMax fields. Since one can solve for the magnetic charge from Eq.~\eqref{Riegertconstraint} and $b$ in terms of $r_+$ from $f(r_+)=0$, this action depends on three parameters in phase space, i.e. $b$, $c$, and $q$. As our main focus in this work is to study gravitational instantons in Conformal Gravity, we leave further thermodynamic explorations of the nonlinearly charged Riegert black hole for a future work.

\subsection{Eguchi-Hanson}\label{sec:Eguchi-Hanson}

In Ref.~\cite{Eguchi:1978xp}, Eguchi and Hanson constructed an asymptotically flat self-dual gravitational instanton in General Relativity, following a similar strategy of Belavin-Polyakov-Schwartz-Tyupkin in Yang-Mills theory~\cite{Belavin:1975fg}. Then, generalizations of their solution were obtained by including the cosmological constant, Maxwell fields~\cite{Eguchi:1978gw}, and also acceleration~\cite{Chng:2006gh}. In higher-curvature gravity, extensions of the Eguchi-Hanson metric have been studied in Refs.~\cite{Wong:2011aa,Hendi:2012zg,Corral:2021xsu,Corral:2022udb,Corral:2025yvr,Fenwick:2025fgg}. Additionally, it has been used as a seed metric to construct higher-dimensional gravitational solitons by oxidizing the original four-dimensional Eguchi-Hanson solution~\cite{Clarkson:2005qx,Clarkson:2006zk,Durgut:2022xzw}. In the presence of conformally-coupled scalars, a backreacted Eguchi-Hanson-like solution was found in Ref.~\cite{Barrientos:2022yoz}. Here, we generalize the vacuum Eguchi-Hanson solution in Conformal Gravity obtained in Ref.~\cite{Corral:2021xsu} by coupling the theory to spin-0 and spin-1 nonlinear conformal matter. 

To this end, we consider an ansatz for the line element, the nonlinear $U(1)$ gauge potential, and the scalar field as
\begin{subequations}\label{EHansatz}
    \begin{align}
    \dd s^2&=\frac{\dd r^2}{f(r)}+\frac{r^2f(r)}{4}(\dd\tau+\cos\vartheta\dd\varphi)^2+\frac{r^2}{4}(\dd\vartheta^2+\sin^2\vartheta\dd\varphi)\,, \\
    A&=a(r)(\dd\tau+\cos\vartheta\dd\varphi)\,, \qquad \phi = \phi(r)\,,
\end{align}
\end{subequations}
respectively. In the absence of matter fields, the three-parameter family of Eguchi-Hanson-like instantons in Conformal Gravity was studied in Ref.~\cite{Corral:2021xsu}, whose line element is given by
\begin{equation}\label{EH-vac}
	f(r)=1+\frac{c}{r^2}+\frac{r^2}{\ell^2}-\frac{a^4}{r^4}+br^4\,,
\end{equation}
subject to the condition $c=-4a^4b\ell^2$. This metric becomes self-dual when $c=b=0$ and the limit $\ell\to\infty$ is taken, thus recovering the standard asymptotically flat Eguchi-Hanson metric in general relativity~\cite{Eguchi:1978xp,Eguchi:1978gw}. Regularity of the solution imposes the condition $b=0$, since the solution becomes singular as $r\to\infty$ of $b\neq0$; this can be seen by computing the Ricci scalar of this solution, that is,
\begin{align}
    R = -\frac{24}{\ell^2} - 48br^2
\end{align}
From the condition on the integration constants below Eq.~\eqref{EH-vac}, one can see that setting $b=0$ might imply that $c=0$, which is an Einstein space. However, by taking the limit $\ell\to\infty$ in Eq.~\eqref{EH-vac} and then setting $b=0$, while keeping $c$ fixed,  yields~\cite{Corral:2021xsu}
\begin{equation}
	f(r)=1+\frac{c}{r^2}-\frac{a^4}{r^4}\,.
\end{equation}
This metric is anti-self-dual, and it saturates the BPS bound of Conformal Gravity. In fact, notice that this space is not conformally Ricci-flat, as can be verified from the Dunajski-Tod theorem by computing the conditions in Eq.~\eqref{constraints-confE}, which lead to $\mathcal{P}=0$ and
\begin{equation}
	\mathcal{T}_{\mu\nu}=-\frac{c(c^2+4a^4)}{(2a^4-cr^2)^2}g_{\mu\nu}\,.
\end{equation}

Let us consider now the backreaction of nonlinear conformal matter on this space. Thus, inserting these ans\"atze into the field equation~\eqref{eom}, we find that the solution of the system is given by
\begin{align}\label{EHsol}
    f(r)=1+\frac{c}{r^2}+\frac{r^2}{\ell^2}+br^4\,,\qquad a(r)=pr^{2e^\gamma}+qr^{-2e^\gamma}\,, \qquad \phi(r)=\frac{1}{r^2}\sqrt{\frac{c}{\nu}}\,, 
\end{align}
where $b,c,\ell,q,$ and $p$ are integration constants, which are related via the constraint
\begin{equation}\label{constraint-EH}
    96\ell^2\nu pqe^\gamma-c(1+192\alpha\nu)=0\,.
\end{equation}
In the case where $c=0$, the space becomes globally self-dual. However, this choice trivializes the scalar field, and it forces the integration constants of the ModMax fields to vanish in order to satisfy the constraint in Eq.~\eqref{constraint-EH}. On the other hand, in the limit $b\to0$ one recovers the metric function reported in Ref.~\cite{Barrientos:2022yoz}, which is globally anti-self-dual in the limit $\ell \to \infty$, as can be seen from Eq.~\eqref{WWplus}. Similar to the vacuum case, the limit $b\to0$ should not be taken just for the sake of comparison, since their curvature invariants are singular as $r\to\infty$. Therefore, in this case, the condition $b=0$ is required for regularity. 

The absence of conical singularities at the horizon $f(r_+)=0$ of the Eguchi-Hanson instanton~\eqref{EHsol} is achieved by identifying the Euclidean time by $\tau\sim\tau+\beta$, whose period is 
\begin{equation}
    \beta=\eval{\frac{8\pi}{rf'(r)}}_{r=r_+}=-\frac{4\pi r_+^2\ell^2}{(2r_+^2+3c)\ell^2+r_+^4} \,.
\end{equation}
The Euler characteristic is $\chi(\mathcal{M})=2$, similar to the Taub-Bolt solution.

There are special limits where this solution becomes (anti)self-dual. To verify this, we compute the following invariants
\begin{subequations}
\begin{align}
\left(W^{\mu\nu}_{\lambda\rho} - \tilde{W}^{\mu\nu}_{\lambda\rho}\right)\left(W_{\mu\nu}^{\lambda\rho} - \tilde{W}_{\mu\nu}^{\lambda\rho}\right)&=\frac{384c^2}{r^8}\label{WWminus}\,,\\
	\left(W^{\mu\nu}_{\lambda\rho} + \tilde{W}^{\mu\nu}_{\lambda\rho}\right)\left(W_{\mu\nu}^{\lambda\rho} +  \tilde{W}_{\mu\nu}^{\lambda\rho}\right)&=\frac{384}{\ell^4}\,.\label{WWplus}
\end{align}
\end{subequations}
As we discussed above, in the limit $c\to0$ the original Eguchi-Hanson instanton of Einstein gravity is recovered, which is self-dual in the absence of the cosmological constant. However, from Eq.~\eqref{WWplus}, one can see that this configuration is also globally anti-self-dual in the limit $\ell\to\infty$.   

Since this gravitational instanton is not asymptotically locally AdS, the renormalization of the Euclidean on-shell action is not guaranteed by conformal invariance of the theory. However, according to the evidence provided in Ref.~\cite{Barrientos:2022yoz}, the renormalization of conserved charges on Eguchi-Hanson spaces can benefit from the introduction of topological terms of the Pontryagin class; the latter are conformal invariants, so they do not spoil the symmetries of the full theory. Adding the $SO(4)$ and $U(1)$ Pontryagin densities, together with an appropriate choice of their coupling constants, we find that the renormalized Euclidean action is given by
\begin{align}\notag
	I^{(\rm ren)} &= I +\left(\frac{1}{144\nu}-\alpha\right)\int_{\mathcal{M}}\dd^4x\sqrt{|g|}\tilde{R}_{\mu\nu}^{\lambda\rho}R^{\mu\nu}_{\lambda\rho} + \frac{1}{4}\int_\mathcal{M}\dd^4x\sqrt{|g|}\tilde{F}_{\mu\nu}F^{\mu\nu} \\
    &= 4\pi\beta\left(\frac{c^2(5-576\nu\alpha)}{96\nu r_+^4}+q^2r_+^{-4e^\gamma}\right)\,,
\end{align}
where $I$ is given in Eq.~\eqref{Itotal}.

The conserved charges of ModMax fields defined via the Gauss law in Eq.~\eqref{QeQm} are modified by the presence of the $U(1)$ Pontryagin density, and they behave as
\begin{equation}
	Q_e=-\frac{1}{4\pi}\int_{\Sigma_\infty}\left(\star P- F\right)=\lim_{r\to \infty}(-2qr^{-2e^\gamma})=0\,.
\end{equation}
Therefore, the electric charge of the ModMax solution is zero in the presence of the Pontryagin density. Additionally, the magnetic charge is divergent for $p\neq0$; this can be seen from Eq.~ \eqref{QeQm}, which gives
\begin{align}
    Q_m&=\lim_{r\to\infty}a(r)=\lim_{r\to\infty} \left(qr^{-2e^\gamma}+pr^{2e^\gamma}\right)\,.
\end{align}
Thus, the regularity of nonlinear conformal fields requires that $p=0$. This condition renders the solution self-dual, and its conserved charges vanish.

\section{Discussion\label{sec:discussion}}

In this work, we studied nonconformally Ricci-flat gravitational instantons in Conformal Gravity, both in vacuum and in the presence of nonlinear conformal matter that also preserves conformal symmetry. To this end, we focused on conformally coupled scalar fields and ModMax electrodynamics. We first revisited the Kerr-NUT-AdS extension constructed in Ref.~\cite{Liu:2012xn}, which possesses a vanishing Bach tensor. Since Conformal Gravity is finite for asymptotically locally AdS spaces, we computed the conserved charges of this solution using the Noether-Wald prescription~\cite{Lee:1990nz,Iyer:1994ys,Wald:1993nt}. In particular, we obtained its mass and angular momentum, which reduce to the corresponding expressions in general relativity in the limit $b\to 0$ and $c\to 0$ of the metric parameters $b,c$. However, other interesting limits are possible. After performing the continuation to Euclidean signature, we explored the curve in parameter space along which the solution becomes globally (anti)-self-dual. Using the Dunajski-Tod theorem~\cite{Dunajski:2013zta}, we showed that the one-parameter extension of the Kerr metric is not conformally Ricci-flat. Its Euclidean on-shell Euclidean action was obtained, which turns out to be proportional to the Chern-Pontryagin index, thereby saturating a gravitational BPS bound. We also demonstrated that this (anti)-self-dual instanton dominates the path integral over the complete projective space $\mathbb{CP}^2$, previously studied by Strominger, Horowitz, and Perry in Ref.~\cite{Strominger:1984zy}.

In the second part of the paper, we found gravitational instantons in Conformal Gravity with conformally-coupled scalars and ModMax fields. In particular, we presented the generalization of two well-known gravitational instantons of general relativity: the Taub-NUT-AdS and Eguchi-Hanson metrics. For both cases, their temperature, the curve in parameter space where they become (anti)-self-dual, their partition function, and nonlinear Abelian charges were obtained. Similarly to the Kerr-NUT-AdS extension, we used the Dunajski-Tod theorem~\cite{Dunajski:2013zta} to show that the vacuum Taub-NUT-AdS instanton in Conformal Gravity is not conformally Ricci-flat. Its Euclidean on-shell action was computed following the prescription in Ref.~\cite{Anastasiou:2022wjq}, where suitable boundary terms are added without breaking the conformal invariance of the action. The electric and magnetic charges associated with the ModMax fields, defined through a Gauss law and through the integral of the first Chern class, respectively, were also computed. We showed that these charges are directly related to the two integration constants appearing in the ModMax field configuration. Furthermore, we analyzed the static limit of the NUT charge $n\to 0$, in which the solution reduces to the massless limit of the Riegert metric~\cite{Riegert:1984zz} dressed with nonlinear conformal matter.

There are several possible extensions of this work. One direction would be to study the holographic properties of the Kerr-NUT-AdS metric extension with weakened boundary conditions. Since Conformal Gravity possesses a well-defined variational principle for holographic sources~\cite{Grumiller:2013mxa}, it would be possible to compute the partially massless response associated with this space, thereby allowing the exploration of holographic features beyond those typically found in standard general relativity. Regarding the gravitational instantons dressed with nonlinear conformal matter presented in this work, another natural extension would be to investigate whether phase transitions to a holographic state exist, in a manner similar to the studies in Refs.~\cite{Hartnoll:2008vx,Hartnoll:2008kx}. These questions are left open for future work.

\begin{acknowledgments}
    The authors thank Eloy Ayón-Beato, Adolfo Cisterna, and Daniel Flores-Alfonso for useful discussions. C.C. thanks the National Technical University of Athens for its hospitality during the initial stage of this work. This work is partially supported by Agencia Nacional de Investigación y Desarrollo, Chile, via Fondecyt Regular grants 1240043, 1240048, 1251523, 1252053, and 1230112.
\end{acknowledgments}

\bibliography{References}

\end{document}